\documentclass[twocolumn,preprintnumbers,floatfix,letterpaper,prc]{revtex4}

\usepackage{citesort,enumerate}
\usepackage{amsmath,amssymb}  
\usepackage{bm}               
\usepackage{amscd}            
\usepackage{graphicx}         
\usepackage{hhline,multirow}  
\usepackage{dcolumn}          
                              
\newcommand{\beq}{\begin{equation}}
\newcommand{\eeq}{\end{equation}}
\newcommand{\bea}{\begin{eqnarray}}
\newcommand{\beas}{\begin{eqnarray*}}
\newcommand{\beau}[1]{\begin{equation} \label{#1} \begin{array}{rcl}}
\newcommand{\eea}{\end{eqnarray}}
\newcommand{\eeas}{\end{eqnarray*}}
\newcommand{\eeau}{\end{array} \end{equation}}
\newcommand{\bay}{\begin{array}}
\newcommand{\eay}{\end{array}}
\newcommand{\btab}{\begin{tabular}}
\newcommand{\etab}{\end{tabular}}
\newcommand{\bals}{\begin{align*}}
\newcommand{\eals}{\end{align*}}

\newcommand{\ra}{{\rightarrow}}

\newcommand{\vev}[1]{\langle #1 \rangle}

\newcommand{\hermes}{{\sc HERMES}}
\newcommand{\clas}{{\sc CLAS}}

\begin{document}


\preprint{JLAB-THY-08-864}

\title{An estimate of the prehadron production time}

\author{Alberto~Accardi$^{a,b}$}
\affiliation{
$^a$Hampton University, Hampton, VA, 23668, USA \\
$^b$Jefferson Lab, Newport News, VA 23606, USA \\
}

\begin{abstract}
A semi-quantitative estimate of the prehadron production time based
on recent preliminary \hermes\ data on hadron transverse momentum
broadening in nuclear DIS is presented. 
The obtained production time can well explain the data except for their
dependence on the photon virtuality, which remains a challenge to
current theoretical models. A few mechanisms that may
contribute to its explanation are suggested, along with possible
experimental tests. 
The average time scale at \hermes\ is found to be of the order of 4-5 fm, 
comparable to heavy nuclei radii, so that by suitable kinematic cuts
it will be possible to study both the case in which hadronization starts
inside and the case in which it starts outside the nucleus.
A comparison with recent CLAS preliminary data, which qualitatively but
not quantitatively agree with HERMES, is performed. 
\end{abstract}



\maketitle


\section{Introduction}

Nuclear modifications of hadron production in high-energy collisions 
have been observed in both Deep Inelastic lepton-nucleus Scattering (nDIS)
\cite{Airapetian:2007vu} and in heavy ion collisions
\cite{Arsene:2004fa,Back:2004je,Adams:2005dq,Adcox:2004mh}. 
One typically observes: (i) a suppression of hadron 
multiplicities, called hadron quenching or jet quenching;
(ii) hadron transverse momentum ($p_T$) broadening; (iii) the related 
modification of the hadron $p_T$-spectrum also known as Cronin
effect. 
The nuclear modifications can be attributed to the
interactions of the scattered partons and of the hadrons formed in
their fragmentation with the surrounding medium. Experimentally,
partonic in-medium interactions can be isolated by studying 
Drell-Yan (DY) lepton pair production in hadron-nucleus
collisions. In nDIS and hadron-nucleus collisions, the medium is the
nuclear target 
itself, also called ``cold nuclear matter''. In nucleus-nucleus
collisions, the fragmenting parton must also traverse the
hot and dense medium created in the collisions, be it a hadron gas at
low energy, or a Quark-Gluon Plasma (QGP) at high
energy. This medium is also called ``hot nuclear matter''.

A precise knowledge of parton propagation and hadronization
mechanisms obtained from nDIS and DY data is essential for testing and
calibrating our theoretical tools, and to determine the properties of
the QGP produced at the Relativistic Heavy Ion Collider. Conversely, a
well known nuclear medium like 
a target nucleus allows testing the hadronization
mechanism and color confinement dynamics in nDIS. Knowledge of
partonic in-medium propagation gained from nDIS, can be used in DY
scatterings to factor out parton energy loss and measure the nuclear
modifications of parton distributions in the initial state
\cite{Garvey:2002sn}. 
Finally, hadron quenching is an important source of systematic
uncertainty in neutrino oscillation experiments such as MINOS, 
which use nuclear targets and need to reconstruct the
event's kinematic from the hadronic final state
\cite{Adamson:2008zt,Dytman:2008st}. 

\begin{figure}[b]
  \centering
  \includegraphics[width=\linewidth]{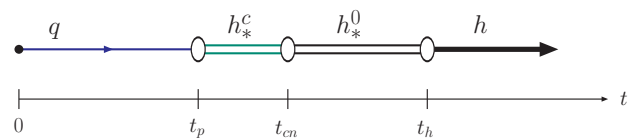}
  \caption{Sketch of the time evolution of the hadronization process
    with definition of the relevant time scales. A quark $q$ created at
    time 0 in a hard collision turns into a colored prehadron $h_*^c$,
    which subsequently neutralizes its color, $h_*^0$, and collapses on the
    wave function of the observed hadron $h$.}
  \label{fig:hadrosketch}
\end{figure}

Understanding and modeling nuclear modifications of hadron production
requires knowledge of the space-time evolution of the hadronization
process \cite{Accardi:2006ea}.
However, hadronization is a non perturbative process, and its theoretical
understanding is still in its infancy: one has to resort to
phenomenological models to describe its space-time evolution
\cite{Bialas:1986cf,Accardi:2002tv,Accardi:2005jd,Kopeliovich:2003py,Falter:2004uc,Gallmeister:2005ad,Gallmeister:2007an}. 
Nonetheless, a few features can be expected on general grounds.
A parton created in a high-energy collision can travel 
as a free particle only for a limited time because of color
confinement: it has to dress-up in a color-field of loosely bound
partons, which eventually will evolve into the observed hadron 
wave function.  
While the naked, asymptotically free, parton has a negligible
inelastic cross section with the surrounding medium constituents, the
dressed parton is 
likely to develop an inelastic cross section of the order of the
hadronic one. Hence, the dressed parton will be subject to nuclear
absorption in a similar way a fully formed hadron is. For this
reason, it is usually called ``prehadron'', and denoted by $h_*$. 
The prehadron may for some time be in a colored state,
$h_*^c$, because color neutrality is only required for the 
final state hadron. However, it is likely to neutralize its color
before hadron formation, and the colorless prehadron is denoted
$h_*^0$. We can therefore identify 3 relevant time 
scales, see Fig.~\ref{fig:hadrosketch}: 
(1) the ``prehadron production time'' or ``quark lifetime''
$t_p$, at which the dressed quark develops a sizable inelastic cross section,
(2) the ``color neutralization time'' $t_{cn}$, at which gluon
bremsstrahlung stops, and (3) the ``hadron formation time'' $t_h$, at
which the final hadron is formed.

For practical applications, hadronization is generally
pictured as a 2 step process in which the prehadron production time 
and color neutralization time are identified, $t_p=t_{cn}$.
This is a somewhat crude approximation of the more complex
process sketched above, but is adequate to the present status of the
theoretical and experimental investigation and will be used in this
paper. 

In summary, the key quantity we need to investigate in order to
understand this complex dynamics is the hadronization time scale,
$t_p$. It is the most general
information about the space-time evolution of the hadronization
process which can be extracted from experimental data. 
It is the purpose of this paper to
provide a semi-quantitative and model independent 
estimate of the prehadron production time
from recent preliminary data on the $p_T$-broadening taken by the
\hermes\ experiment in lepton-nucleus scatterings at the HERA
accelerator
\cite{VanHaarlem:2007kj,Jgoun-prelim,vanHaarlem:2007zz}. 
These data show rich 
features in terms of their dependence on the kinematic variables,
which are confirmed by preliminary data taken by the \clas\ experiment
at Jefferson Lab \cite{Brooks-Trento,Hicks-Trento,Hafidi:2006ig}, 
and are not always simple to interpret theoretically. 
I will show to what extent a consistent picture of the space-time
evolution of hadronization can be inferred from these data, and how
they are beginning to expose the limits of current theoretical
modeling of this process.

\section{Production time scaling}
\label{sec:scaling}

Hadron quenching in nDIS is studied in terms of the multiplicity ratio
\begin{equation} 
  R_M^{h}(z,\nu, p_t^2, Q^2) = {\frac {\left. \frac{N_h(z,\nu, p_t^2,
          Q^2)}{N_e(\nu, Q^2)}\right|_A } 
   {\left. \frac{N_h(z,\nu, p_t^2, Q^2)}{N_e(\nu, Q^2)}\right|_D }},
\label{eq:att}
\end{equation}
where  $N_h$ is the yield of semi-inclusive hadrons in a given
kinematic bin, and $N_e$ the yield of inclusive scattered
leptons in the same ($\nu$,$Q^2$)-bin.
The kinematic variables $z$, $\nu$, $p_t^2$, $Q^2$ are the usual DIS
invariants, namely, the hadron fractional energy,
the virtual photon energy, the hadron transverse momentum and the
lepton 4-momentum transfer squared.

In Ref.~\cite{Accardi:2006qs} it is conjectured that $R_M$ should not
depend on $z$ and $\nu$ separately but should
depend on a combination of them:
\begin{align}
  R_M = R_M\big[\tau(z,\nu)\big] \ ,
 \label{eq:RMscaling}
\end{align}
where the scaling variable $\tau$ is defined as
\begin{align}
  \tau & = C\, z^\lambda (1-z) \nu \ .
 \label{eq:scalingvar}
\end{align}
The scaling exponent $\lambda$ can be obtained by a best fit analysis
of data or theoretical computations. Obviously, the proportionality
constant $C$ cannot be determined by the fit. 
A possible scaling of $R_M$ with $Q^2$ is not considered 
because of its model dependence, and because of 
the mild dependence of HERMES $R_M$ data on $Q^2$.
As discussed below, the proposed functional
form of $\tau$, Eq.~\eqref{eq:scalingvar}, is flexible enough to
encompass both absorption models, which assume short production times
and in-medium hadronization, and energy loss models, which assume long
lived quarks with $\vev{t_p} \gg R_A$, where $R_A$ is the nuclear
radius. The 2 classes of models are distinguished by 
the value of the scaling exponent: a positive $\lambda \gneqq 0$ is
characteristic of absorption models, while a negative $\lambda \lesssim
0$ is characteristic of energy loss models. Thus, the exponent
$\lambda$ extracted from experimental data can identify 
the leading mechanism for hadron suppression in
nDIS, and distinguish short from long hadronization time scales.

The scaling of $R_M$ is quite natural in the context of hadron
absorption models
\cite{Bialas:1986cf,Accardi:2002tv,Accardi:2005jd,Kopeliovich:2003py,Falter:2004uc,Gallmeister:2005ad}. 
Indeed, prehadron absorption depends on the
in-medium prehadron path length, which depends on
the prehadron production time $\vev{t_p}$, as long as $\vev{t_p}
\lesssim R_A$. 
In the Lund string model \cite{Andersson:1983ia} hadronization is
modeled by the breaking of the color string stretching from the struck
parton to the target remnant. The production time is    
\begin{align}
  \vev{t_p} & = f(z) (1-z) \frac{z E_q}{\kappa_\text{str}} 
 \label{eq:lundest}
\end{align}
where $E_q$ is the struck quark energy, and $\kappa_\text{str}$ the
string tension. At leading order (LO) in the Strong coupling constant
$\alpha_s$, the partonic subprocess is $\gamma^*+q \ra q$ and one obtains
\begin{align}
  E_q = \nu \ .
\end{align}
The factor $z E_q$ can be understood as a Lorentz boost factor.
The $(1-z)$ factor is due to energy conservation: a high-$z$ hadron
carries away an energy $zE_q$; the string remainder has a small
energy $\epsilon=(1-z)E_q$ and cannot stretch farther than
$L=\epsilon/\kappa_{str}$. Thus the string breaking must occur on a
time scale proportional to $1-z$. The function $f(z)$ 
is a small deformation of $\vev{t_p}$, which can be computed
analytically in the standard Lund model
\cite{Accardi:2005jd,Bialas:1986cf}.  
The main features of the estimate \eqref{eq:lundest} are dictated by
kinematics and 4-momentum conservation, hence are of general
nature. Indeed they can be obtained also by perturbative considerations
based on the uncertainty principle, similarly to what is discussed in
Ref.~\cite{Adil:2006ra}, or in the perturbative hadronization model of
Ref.~\cite{Kopeliovich:2003py}. 
The production time \eqref{eq:lundest} is well described by the
proposed scaling variable $\tau$ with $\lambda>0$. E.g., in the  Lund
model $\lambda \approx 0.7$ \cite{Accardi:2006qs}.  
In energy loss models \cite{Wang:2002ri,Arleo:2003jz,Accardi:2005mm},
which assume $\vev{t_p} \gg R_A$,
the scaling is less obvious and holds only approximately on a theoretical
ground. When performing the scaling analysis of the full energy loss
models, one finds in general $\lambda \lesssim 0$
\cite{Accardi:2006qs}.  

The central result of the analysis of HERMES data at $E_\text{lab}=27$
GeV performed in Ref.~\cite{Accardi:2006qs} is
that pion data clearly exhibit
\begin{align}
  \lambda \approx 0.5 \gneqq 0 \ .
\end{align}
As discussed, this 
is a signal of in-medium prehadron formation, with production times 
$\vev{t_p} = O(R_A)$. Therefore, the scaling variable $\tau$ can be
identified with the average production time:
\begin{align}
  \vev{t_p} \equiv \tau \ .
\end{align}
The proportionality constant $C$ in Eq.~\eqref{eq:scalingvar}, 
hence the magnitude of the production time, will be estimated in
Section~\ref{sec:discussion-ptbroad}.  
A similar analysis has been attempted in Ref.~\cite{Airapetian:2007vu},
but arbitrarily fixing $\lambda = 0.35$. The resulting $R_M(\tau)$
shows a rough scaling, with the 2 lowest-$z$ data point in each data
set deviating from the scaling curve and aligning in the vertical
direction. This breaking of the scaling behavior is more probably 
due to the particular choice of $\lambda$, rather than to rescattering
effects as argued by the authors.

\section{$p_T$-broadening and prehadron formation} 
\label{sec:discussion-ptbroad}

\begin{figure}[t]
  \centering
  \parbox[b]{0.50\linewidth}{
    \includegraphics[width=\linewidth]{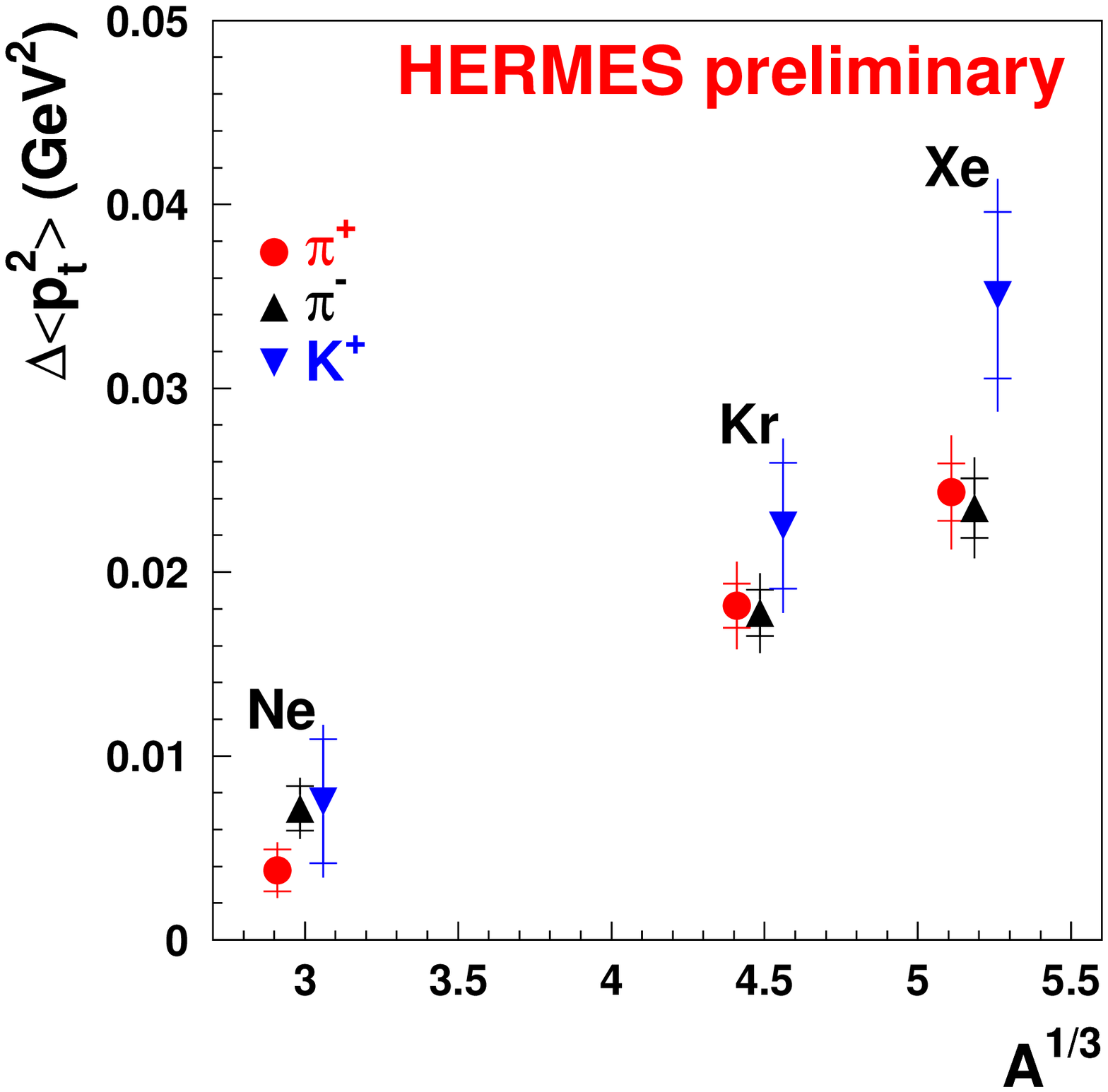}
  }
  \parbox[b]{0.47\linewidth}{
    \includegraphics[width=\linewidth]{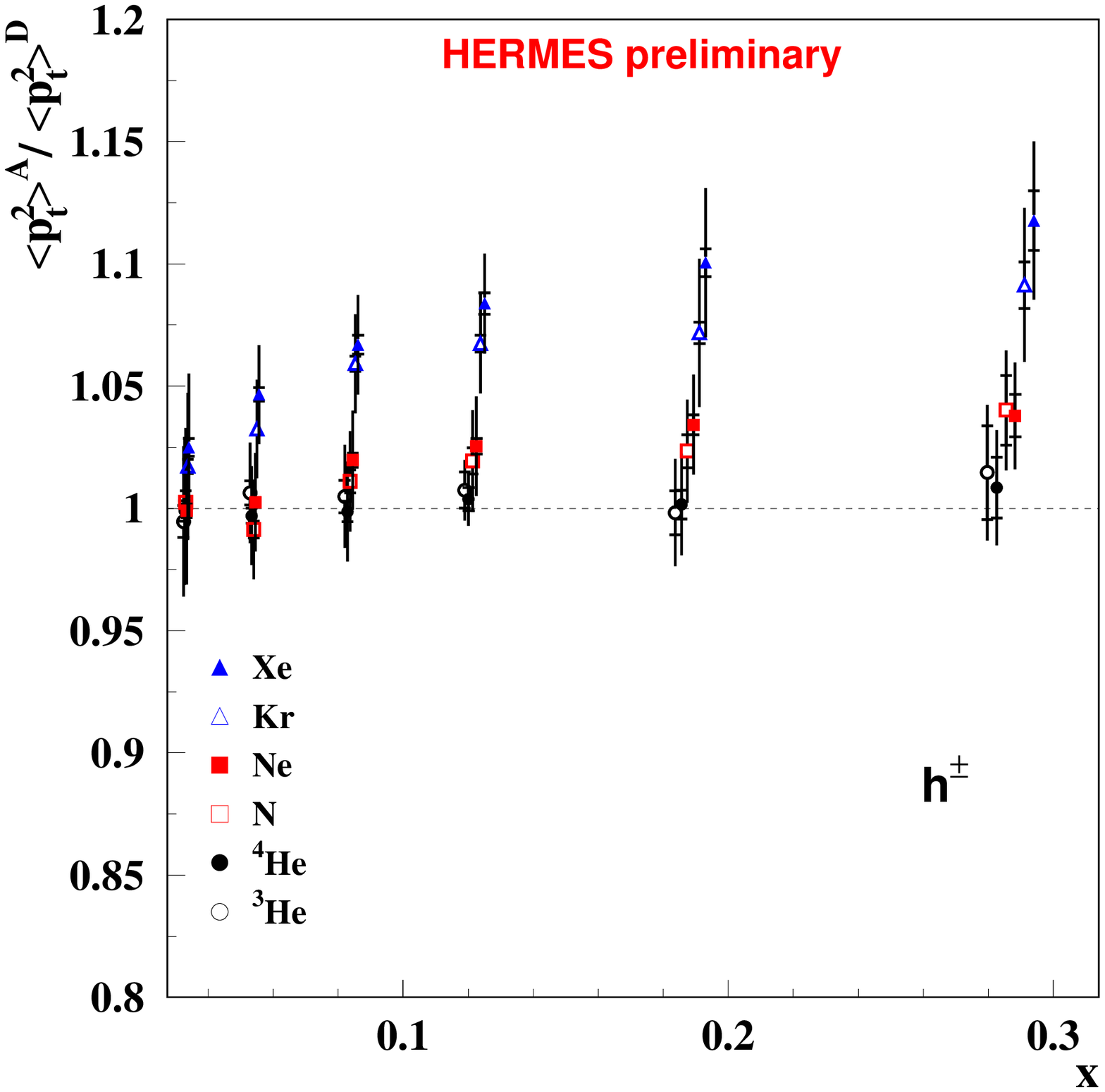}
    \vskip0.247cm
  }
  \caption{Left: $p_t$-broadening at \hermes\ for different hadron types
    produced from several nuclear targets as a function of the atomic
    number $A$ (from Refs.~\cite{vanHaarlem:2007zz,VanHaarlem:2007kj}).
    Right: Ratio $\vev{p_T^2}_A/\vev{p_T^2}_D$ as a function of $x_B$
    (from Ref.~\cite{Jgoun-prelim}). The inner 
    error bars represent the statistical error and the outer ones the
    quadratic sum of the statistical and systematic
    uncertainties.
  }
  \label{fig:hermesptbroad_A}
\end{figure}

The scaling analysis just described 
gives only indirect evidence for a short production time $\vev{t_p}$,
and cannot measure its absolute scale. 
An observable which is more
directly related to the prehadron production time, and allows an
estimate of the coefficient $C$ in Eq.~\eqref{eq:scalingvar}, is the 
hadron's transverse momentum broadening in DIS on a nuclear target
compared to a proton or deuteron target
\cite{Kopeliovich:2003py,Kopeliovich:2006xy},
\begin{align}
  \Delta\vev{p_T^2} = \vev{p_T^2}_A - \vev{p_T^2}_D \ .
\end{align}
When a hadron is observed in the final state, neither the
quark nor the prehadron from which it originates could have had 
inelastic scatterings. Since the prehadron-nucleon elastic 
cross section is very small compared to the quark cross section, 
the hadron's $p_T$-broadening originates dominantly during parton
propagation. As shown in
\cite{Baier:1996sk,Johnson:2000dm}, the quark's momentum
broadening $\Delta\vev{p_T^2}$ is proportional to the quark path-length
in the nucleus. If the prehadron production time 
has the form \eqref{eq:scalingvar} as argued in the last section, and
as long as the prehadron is formed inside the nucleus, we obtain 
\begin{align}
  \Delta \vev{p_T^2} \propto \vev{t_p} \propto z^{0.5} (1-z) \nu  \ ,
  \label{eq:ptvsznu}
\end{align}
where the exponent 0.5 is determined by the scaling analysis discussed
in Section \ref{sec:scaling}.
Then, a decrease of $\Delta p_T^2$ with increasing
$z$ or decreasing $\nu$ would be a clear signal of in-medium prehadron
formation: indeed, if the quark were traveling
through the whole nucleus before prehadron formation, 
$\Delta p_T^2$ would only depend on the nucleus size and not on
$z$ or $\nu$. 
In addition to the dependence of $\Delta\vev{p_T^2}$ on $z$ and
$\nu$, which primarily 
stems from energy conservation and the Lorentz boost of the hadron,
Ref.~\cite{Kopeliovich:2003py} argues that 
\begin{align}
  \Delta \vev{p_T^2} \propto \frac{1}{Q^2} \ .
  \label{eq:ptvsq2}
\end{align}
The physics behind this proposed behavior is that a quark which is
struck by a photon of large virtuality radiates more intensely than
for a lower virtuality: as a consequence, it will be able to only
travel a shorter way before hadronization, hence it will experience less
$p_T$-broadening.
A related observable is the Cronin effect, which is likewise expected
to decrease with increasing $z$ 
or decreasing $\nu$, and possibly with increasing $Q^2$
\cite{Kopeliovich:2003py}.

\begin{figure}[t] 
  \centering
  \includegraphics[width=1\linewidth]{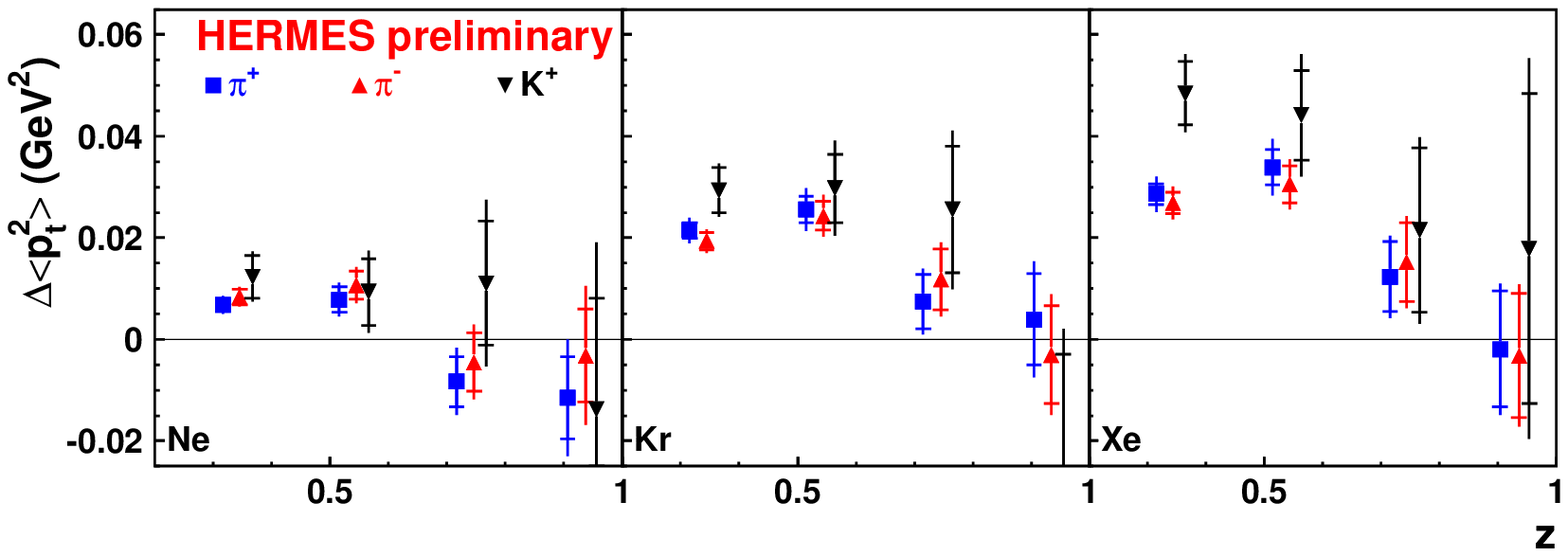}
  \includegraphics[width=1\linewidth]{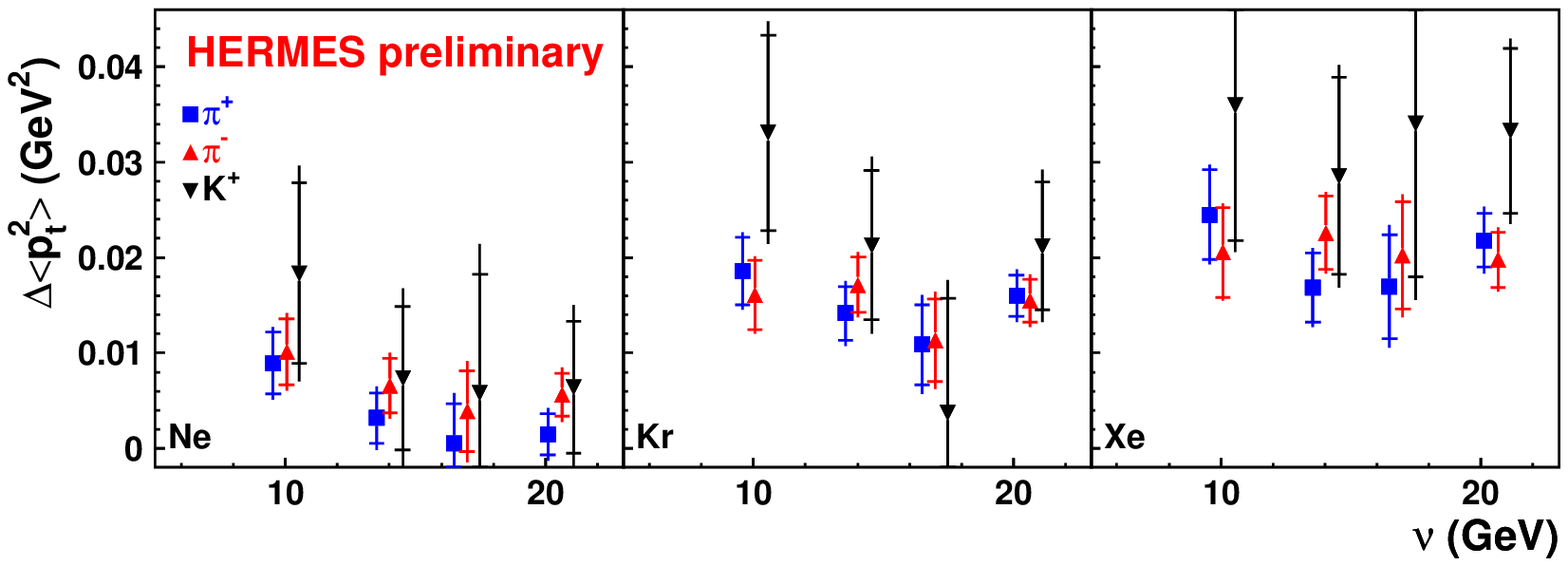}
  \includegraphics[width=1\linewidth]{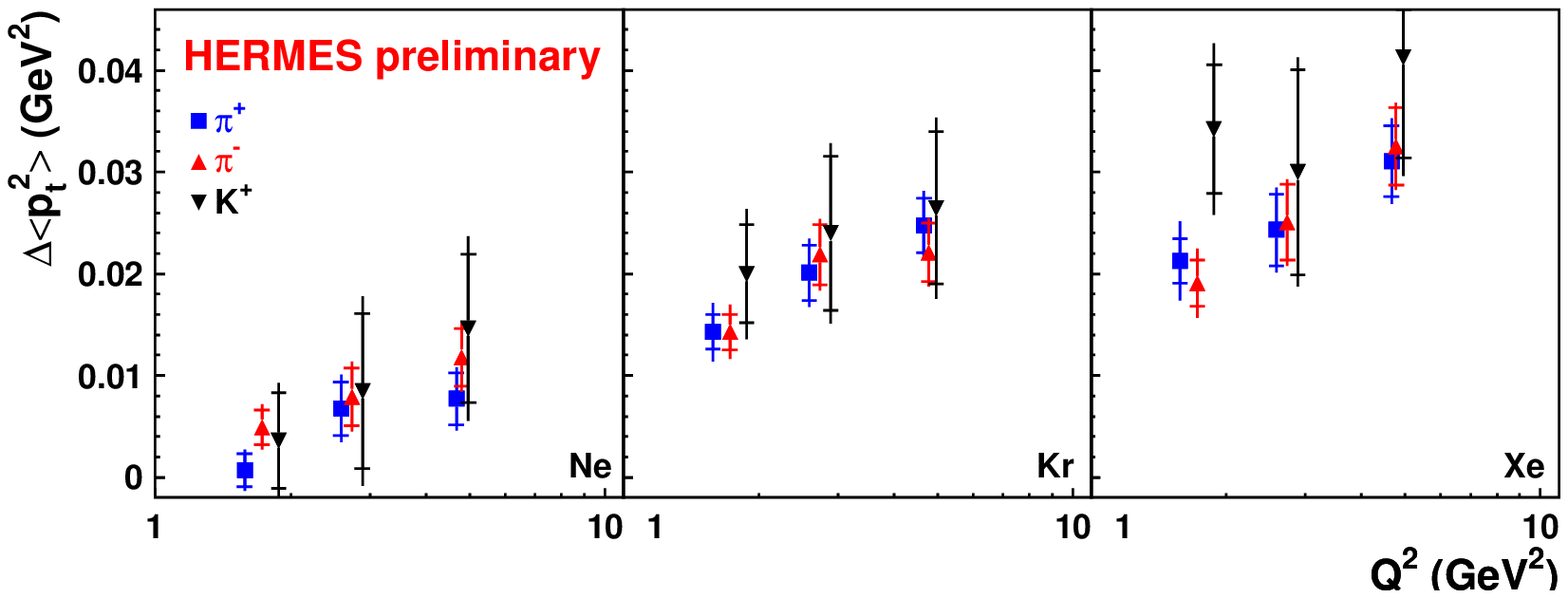}
  \caption{As Figure~\ref{fig:hermesptbroad_A}, but for the 
    $p_t$-broadening as a function of $z$ (upper
    panel), $\nu$ (middle panel), and Q$^2$ (lower panel). Plots taken
    from Refs.~\cite{VanHaarlem:2007kj,vanHaarlem:2007zz}. 
  }
  \label{fig:hermesptbroad_znuQ2}
\end{figure}

The HERMES preliminary data on the $p_T$-broadening, 
reproduced in Figs.~\ref{fig:hermesptbroad_A} and
\ref{fig:hermesptbroad_znuQ2} for the reader's convenience, show 
a linear increase with $A^{1/3}$, and a clear
decrease of $\Delta\vev{p_T^2}$ at large $z$ consistent with
Eq.~\eqref{eq:ptvsznu} and HERMES data on the Cronin effect
\cite{Airapetian:2007vu}. 
However, HERMES data also show a
clearly increasing $\Delta \vev{p_T^2}$ with $Q^2 = 2-10$ GeV$^2$,
at variance with the $Q^2$ dependence of the prehadron production time
\eqref{eq:ptvsq2} obtained in Ref.~\cite{Kopeliovich:2003py}, and with
string models in general. They also show an increase of $\vev{p_T^2}|A
/ \vev{p_T^2}|D$ with $x_B$ tending to saturate at $x_B \gtrsim 0.2$
\cite{Jgoun-prelim}. The flat, or even slightly decreasing 
dependence of $\Delta\vev{p_T^2}$ on $\nu$ is also, at first sight, 
not easily interpreted in the context of Eq.~\eqref{eq:ptvsznu}
and the simple hadronization picture used so far. 

\begin{table}[tb]
\centering
\begin{tabular}{|c||c|c|c|c|c|}
  \hline
    & $\vev{Q^2}$ 
    & $\vev{\nu}$ 
    & $\vev{z}$
    & $\frac{\vev{Q^2}}{2m_N\vev{\nu}}$
    & $\vev{t_p}$ \\
    & \footnotesize [GeV$^2$]
    & \footnotesize [GeV]
    & 
    &
    & \footnotesize [fm]\\\hline
  $\vev{\Delta p_{Th}^2}$ vs $A$ & & & & & \\
  Ne (2.3 fm) & 2.4 & 13.7 & 0.42 & 0.09 & 4.2 \\
  Kr (3.7 fm) & 2.4 & 13.9 & 0.41 & 0.09 & 4.2 \\
  Xe (4.3 fm) & 2.4 & 14.0 & 0.41 & 0.09 & 4.3 \\\hline
  $\vev{\Delta p_{Th}^2}$ vs $z$ 
     & 2.4 & 14.6 & 0.30 & 0.09 &  4.5 \\
     & 2.4 & 13.3 & 0.53 & 0.10 &  3.7 \\
     & 2.3 & 12.6 & 0.74 & 0.10 &  2.3 \\
     & 2.2 & 10.8 & 0.92 & 0.11 &  0.7 \\\hline
  $\vev{\Delta p_{Th}^2}$ vs $\nu$  
     & 2.1 &  8.1 & 0.48 & 0.14 &  2.4 \\ 
     & 2.5 & 12.0 & 0.42 & 0.11 &  3.7 \\
     & 2.6 & 15.0 & 0.40 & 0.10 &  4.6 \\
     & 2.4 & 18.6 & 0.36 & 0.07 &  5.8 \\\hline
  $\vev{\Delta p_{Th}^2}$ vs $Q^2$ 
     & 1.4 & 14.0 & 0.41 & 0.06 &  4.2 \\
     & 2.4 & 14.1 & 0.41 & 0.10 &  4.2 \\
     & 4.5 & 14.5 & 0.39 & 0.16 &  4.3 \\
  \hline
\end{tabular}
\caption{Average HERMES kinematics for the $p_T$-broadening
  results \cite{vanHaarlem:2007zz,VanHaarlem:2007kj}. 
  In parenthesis, beside the target nucleus symbol is the
  average in-medium path length of the hadronizing system 
  $\vev{L_A}\approx (3/4)R_A$, with $R_A=(1.12 \text{\ fm})A^{1/3}$. The
  production time is computed according to Eq.~\eqref{eq:prodtimeest}.
  The average $\vev{x_B}$ is very well approximated by
  $\vev{Q^2}/(m_N\vev{\nu})$ \cite{VanHaarlem-private}. 
  }
\label{tab:HERMESkin}
\end{table}

\begin{figure}[tb]
  \centering
  \includegraphics[width=0.8\linewidth,trim=0 190 0 0,clip=true]{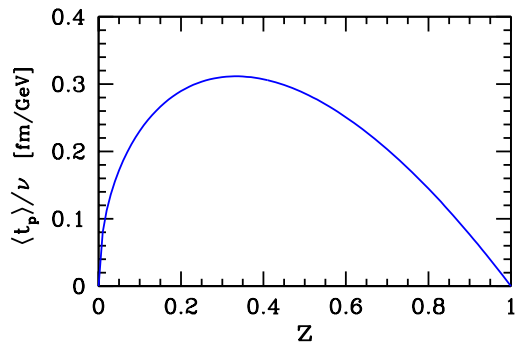}
  \caption{$z$ dependence of the prehadron production time
    $\vev{t_p}/\nu\equiv\tau\nu$ extracted from HERMES data.} 
  \label{fig:prodtime}
\end{figure}

For a more quantitative interpretation of the data we have first 
to specify the average kinematics in each bin, see
Table~\ref{tab:HERMESkin} and Ref.~\cite{vanHaarlem:2007zz}. 
Indeed, different observables and different experimental bins may be
related to different prehadron production times. To this purpose, let
us consider the $A$ and $z$ distributions: 
\begin{itemize}
\item 
The scaling of $\Delta\vev{p_T^2} \propto
A^{1/3}$ up to the Xe nucleus indicates that the quark path length is
larger than the average in-medium path length $\vev{L_\text{Xe}}$ 
of the hadronizing system in the Xe nucleon at $\vev{z} \approx 0.4$,
$\vev{\nu}\approx 14$ GeV, so that the prehadron is always formed on
the surface or outside the nucleus. 
\item
The above kinematics is close to that of the first 2 $z$-bins at
$z=0.30$ and 0.53. Hence we may assume that also in these 2 bins the
prehadron is formed on the surface or outside the nucleus. However,
given the decrease of $\Delta\vev{p_T^2}$ with $z$ as $z\gtrsim 0.5$,
the prehadron must soon enter the nucleus at larger $z$. Hence, the 
production time at $z\approx 0.4$ fm cannot be much larger than
$\vev{L_\text{Xe}}$. 
\end{itemize}
These 2 remarks allow setting the scale for the production time in
Eq.~\eqref{eq:scalingvar}:
\begin{align}
  C \approx \frac{\vev{L_\text{Xe}}}{\bar z^\lambda (1-\bar z)
    \bar\nu} \approx 0.8 \frac{\text{fm}}{\text{GeV}}
  \label{eq:Cest}
\end{align}
with $\bar z=0.4$ and $\bar\nu=14$ GeV. The average in-medium
path-length of the hadronizing system can be approximated as 
$\vev{L_\text{Xe}} \approx (3/4) R_A$ with $R_A = (1.12 
\text{\ fm}) A^{1/3}$, as if nucleons were uniformly
distributed in the nucleus. The resulting production time,
\begin{align}
  \vev{t_p} & \approx  0.8\,  z^{0.5} (1-z) \,\nu \text{\ fm/GeV} \ ,  
 \label{eq:prodtimeest}
\end{align}
with $\nu$ measured in GeV, is plotted in Fig.~\ref{fig:prodtime}. 
In principle, $C$ should be
allowed to depend on $Q^2$ and $x_B$, but one can neglect it in first
approximation at least for the discussion of the $A$-, $z$-, 
where $\vev{Q^2}$ and $\vev{x_B}$ are rather constant.  
With Eq.~\eqref{eq:prodtimeest}, we can compute the 
production times for each experimental bin, see Table~\ref{tab:HERMESkin}. 

The $\nu$-distribution is also consistent with this estimate:
given the production times in Table~\ref{tab:HERMESkin}, 
it is not too surprising that the $\nu$-dependence is basically flat,
because the prehadron is formed on average on the surface or outside
all the studied nuclei. Three effects can contribute to tilt the
curve and producing the slightly decreasing data: 
(i) a possible dependence of the prehadron
cross section with $\nu$, as it happens for the hadron cross section;
(ii) medium modifications of the DGLAP parton shower
\cite{Ceccopieri:2007ek,Domdey:2008gp,Borghini:2005em,Armesto:2007dt},
which can modify the linear dependence of
$\Delta\vev{p_T^2}$ on the energy loss $\Delta E$ found in
\cite{Baier:1996sk}, and implicit in Eq.~\eqref{eq:ptvsznu}; 
(iii) the correlation between the $p_T$-broadening and the average
Bjorken variable, $\vev{x_B}$, whose origin will be discussed in
detail in the next Section, and which may in fact be the dominant
effect.

\section{$\bm p_T$-broadening vs. $\bm Q^2$ and $\bm x_B$}
\label{sec:Q2xBdist}

We are now in the position to address the intriguing $Q^2$ distribution. 
The linear increase of $\Delta\vev{p_T^2}$ with $Q^2$ is in sharp
contrast with the inverse power dependence obtained in the color dipole
model, see Eq.~\eqref{eq:ptvsq2}.
On the other hand also a flat $Q^2$, as would be obtained in most
string-based models, seems at variance with 
these data. To my knowledge, no model predicts a substantial
rise of $\vev{t_p}$ with virtuality: how can we explain these data? 
From the production times shown in
Table~\ref{tab:HERMESkin}, we can see that the prehadron is always formed
on the surface or outside the nucleus, so that the observed
$Q^2$-dependence must predominantly have a partonic origin. It cannot
simply come 
from multiple parton elastic scatterings, because it would be in first
order proportional to the in-medium parton path-length, which is
basically fixed. The following 3 mechanisms may contribute to
explain these data:
\begin{enumerate}
\item 
{\it Medium-enhanced DGLAP evolution.}
The DGLAP evolution in the kinematics under consideration happens
entirely inside the nucleus, because the prehadron is produced at its
surface. The scale $Q_0^2$ at which hadronization
takes place is not expected to strongly depend on the presence of a
medium  
\footnote{The partial deconfinement of bound nucleons in a nucleus
  \cite{Close:1984zn} may induce an $A$-dependence of $Q_0$
  and modify hadron fragmentation \cite{Accardi:2002tv,DiasdeDeus:1985mg}.  
  However, such an effect is unfavored by the measured $Q^2$ dependence of
  $R_M$, which is opposite than what is expected in the partial
  deconfinement mechanism \cite{Accardi-Santorini}. }.
Therefore, in the large-$Q^2$ bins, a longer and 
medium-enhanced DGLAP evolution
\cite{Ceccopieri:2007ek,Domdey:2008gp,Borghini:2005em,Armesto:2007dt}
would imply a larger $p_T$-broadening than at low $Q^2$ . 
How strong this effect is, and whether it can lead to a linear increase of
the $p_T$-broadening with $Q^2$ remains to be seen. 
This mechanism can be investigated with a $\nu > \nu_\text{min}$ cut in
the experimental data: as $\nu_\text{min}$ is increased, 
the DGLAP evolution would increasingly happen outside the nucleus, and
the slope in $Q^2$ should become smaller and smaller. 
\item 
{\it Next-to-leading order processes.}
The basic hard partonic process considered so far is the leading order
in $\alpha_s$ quark elastic
scattering $\gamma^*+q\ra q$. It is the dominant process in the
valence region at large $x_B$. As $x_B$ decreases, however, the gluon
distribution in the nucleon quickly rises, so that the photon-gluon
fusion process $\gamma^*+g \ra q+\bar q$, which is a next-to-leading
order (NLO) process, becomes non negligible. In 
this case the photon energy $\nu$ is shared by the quark and the
antiquark, so that $E_q < \nu$, 
therefore reducing the quark lifetime and its $p_T$-broadening
compared to the LO case where $E_q = \nu$. 
As can be seen from
Table~\ref{tab:HERMESkin}, the average $\vev{x_B}$ spanned in the
$Q^2$ distribution lies in the transition region between sea partons and
valence quark dominance. Hence, the competition between LO and NLO
process might alter the naive picture of $p_T$-broadening adopted so
far, and lead to its increase with $Q^2$. This mechanism may also be
at work in $\nu$-distributions, although with smaller effects because
of the more limited range spanned by $\vev{x_B}$, and contribute to
tilt their slope downwards.
Note that if this mechanism is indeed at work, the estimate of
$C$ presented in Eq.~\eqref{eq:Cest} only gives a lower limit on the
hadron production time. 
Experimentally, this mechanism can be tested by measuring
$Q^2$-distributions with suitable cuts on $x_B$. The slope in $Q^2$
should become flat at very small or large $\vev{x_B}$, see the
discussion below.
\item 
{\it Colored prehadrons with short formation times.}
A more radical possibility is that at these values of
$\vev{\nu}$, the quark does not propagate freely for a long
time; instead, shortly after the hard interaction, it turns into a
colored prehadron $h_*^{c}$ which can loose energy by gluon
bremsstrahlung, thereby broadening its transverse momentum. 
The prehadron may have
an inelastic cross section $\propto 1/Q^2$,
growing in time to the full hadronic one. If the time
evolution is slow enough, at low $Q^2$ the prehadron would be subject
to more nuclear absorption than at large 
$Q^2$, explaining the linear rise of $\Delta\vev{p_T^2}$ with $Q^2$ 
\footnote{A similar space-time evolution of hadronization has been
  implemented for leading prehadrons in the GiBUU absorption model of
  Refs.~\cite{Falter:2004uc,Gallmeister:2007an}, which however 
  assumes $t_p=t_{cn}=0$ for leading (but colorless) prehadrons, 
  and does not address radiative energy loss. It
  would be interesting to phenomenologically include both a colored
  and a colorless prehadron in this model to test the proposed
  scenario.}. 
An extreme version of this mechanism would involve
\begin{align}
\begin{split}
  \vev{t_p} & \approx 0 \\
  \vev{t_{cn}} & \approx  0.8\,  z^{0.5} (1-z) \,\nu \text{\ fm/GeV} \ .
\end{split}
\end{align}
Note that this scenario does not contradict the dipole model
prediction that $\vev{t_p}\propto 1/Q^2$, but would require the prehadron
to propagate as a colored state for a time of the order of the nuclear
radius. An experimental investigation of this scenario needs very
large $\nu$, outside the reach of the HERMES experiment but attainable 
at the Electron-Ion Collider (EIC)
\cite{EIChomepage,Deshpande:2005wd}, in order to significantly boost
$\vev{t_p}$ and allow the quark to propagate as a 
free particle inside the nucleus.  
\end{enumerate}

In order to study the physics behind the experimental correlation of
$\Delta\vev{p_T^2}$ with $x_B$ and $Q^2$, one needs to simultaneously
address $Q^2$-distributions binned in $x_B$ and $x_B$-distributions
binned in $Q^2$. In this way, one can factor out the trivial kinematic
correlation $\nu=Q^2/(2m_Nx_B)$ which affects the production time.
For example, if $\vev{t_p} \propto \nu/\kappa$ with $\kappa$ fixed,
as in the Lund string model,
for the LO $\gamma^*+q\ra q$ scattering we should expect 
\begin{align}
\begin{split}
  Q^{-2}\Delta\vev{p_T^2}_{x_B\text{-bins}} \approx \text{const.} \\
  x_B \Delta\vev{p_T^2}_{Q^2\text{-bins}} \approx \text{const.} 
\end{split}
\end{align}
The deviation from this {\it combined} scaling is going to expose the
underlying dynamics, see Table~\ref{tab:scalingpt2}: if mechanism
number 2 is dominant, we would 
observe $Q^{-2}\Delta\vev{p_T^2}_{x_B}$ constant in $Q^2$ but
$x_B\Delta\vev{p_T^2}_{Q^2}$ increasing with $x_B$; mechanisms number
1 or 3 would produce an increasing $Q^{-2}\Delta\vev{p_T^2}_{x_B}$ but 
a constant $x_B\Delta\vev{p_T^2}_{Q^2}$. In the color dipole
model \cite{Kopeliovich:2003py}, $t_p\propto \nu/Q^2 \propto 1/x_B$
hence $Q^{-2}\Delta\vev{p_T^2}_{x_B}$ would be decreasing and
$x_B\Delta\vev{p_T^2}_{Q^2}$ constant.

\begin{table}[tb]
\centering
\begin{tabular}{|cc|c|c|}
  \hline
  & & $Q^{-2}\Delta\vev{p_T^2}_{x_B}$
    & $x_B \Delta\vev{p_T^2}_{Q^2}$    \\
  \multicolumn{2}{|c|}{model} 
    & vs. $Q^2$
    & vs. $x_B$    \\\hline
  $t_p\propto\nu/\kappa$ 
    & LO                  & $\leftrightarrow$ & $\leftrightarrow$ \\
    & mDGLAP (1)          & $\uparrow$        & $\leftrightarrow$ \\
    & NLO vs. LO (2)      & $\leftrightarrow$ & $\uparrow$        \\
    & colored $h_c^*$ (3) & $\uparrow$        & $\leftrightarrow$ \\
  $t_p\propto\nu/Q^2$ 
    & color dipole \cite{Kopeliovich:2003py} 
                          & $\downarrow$      & $\leftrightarrow$ \\   
  \hline
\end{tabular}
\caption{
  Expected results of the scaling analysis of the $p_T$-broadening for
  the models and mechanisms discussed in this paper. The
  numbers in parenthesis refer to the mechanisms listed in
  Section~\ref{sec:Q2xBdist}. The up and down pointing arrows indicate
  increasing or decreasing $p_T$-broadening, the double horizontal
  arrow an approximately constant behavior.  
  }
\label{tab:scalingpt2}
\end{table}

It is also important to extend as much as possible the range in $x_B$
over which the $p_T$-broadening is measured: at small $x_B\lesssim
0.01$ the NLO photon-gluon fusion is the dominant partonic process,
while at $x_B \gtrsim 0.3-0.4$ the photon dominantly scatters at LO
valence quarks. Hence, with mechanism number 2, one may expect
$\Delta\vev{p_T^2}$ to flat in $Q^2$ in these 2 regions.

\section{Comparison to CLAS preliminary data}

A last question needs to be addressed: how do the HERMES data and
these scenarios compare to preliminary CLAS data
\cite{Hafidi:2006ig,Brooks-Trento,Hicks-Trento} on the
$p_T$-broadening and Cronin effect? 

At \clas, the higher beam luminosity allows multidimensional binning
which is only partly 
accessible at \hermes. The  
kinematics is a bit different due to the lower beam energy,
$E_\text{beam} = 5$ GeV compared to 27 GeV. The main
difference is that $\vev{\nu}=2-5$ GeV is much smaller than at HERMES,
therefore prehadron production should typically happen on shorter time
scales, mainly inside the nucleus according 
to the estimate \eqref{eq:ptvsznu}. Note also that $Q^2 = 1-4$ GeV$^2$, and
that, typically, $\vev{x_B}_\text{CLAS} > \vev{x_B}_\text{HERMES}$.
The main features of CLAS data on C, Fe, and Pb targets
are the following:
\begin{itemize}
\item 
The $p_T$-broadening is linear in $A^{1/3}$ at low $A$, but tends to
saturate for large nuclei.
The flattening is confirmed by the lack of increase in the Cronin
effect from the Fe target to the Pb target.
This shows the prehadron being formed on a
time scale $t_p \gtrsim (4/3) R_\text{Fe} = 4.3$ fm, smaller than at
HERMES. 
\item
The $p_T$-broadening is rising and then saturating with $\nu$. The
saturation appears at $\nu 
\gtrsim 4$ GeV, where the HERMES $p_T$ is also approximately flat. 
\item
The Cronin effect decreases with $z$ as it happens at HERMES.
\item 
The Cronin effect markedly increases as $x_B$ increases from 0.1 to
0.5. This confirms the $x_B$ dependence of \hermes\ $p_T$-broadening.
\end{itemize}
All these features corroborate the discussed HERMES data.
However, the production time I have estimated is about a factor 5
smaller than the production time extracted in 
Ref.~\cite{Kopeliovich:2006xy} from CLAS $\nu$-distributions using the
color dipole formalism \cite{Johnson:2000dm}.
On the other hand, based on this paper's analysis, and taking into
account the uncertainty in the estimate of $\vev{L_A}$, HERMES data can
only support up to a doubling of $t_p$, which would otherwise become
incompatible with the large-$z$ decrease of the $p_T$-broadening. 
The \hermes\ and \clas\ data sets may be qualitatively reconciled 
by taking into account the observed increase of the 
$p_T$-broadening with $\vev{x_B}$ discussed
in the previous Section. Indeed, $\vev{x_B}_\text{CLAS} >
\vev{x_B}_\text{HERMES}$, which implies a larger $p_T$-broadening at
\clas.

The origin of the discrepancy between the production times extracted
from HERMES and CLAS data on the $p_T$-broadening remains to be
clarified. On the theoretical side, the two discussed
methods to determine the production time should be 
applied to both data sets, in order to check their
consistency and further explore the issue. On the experimental side,
it is very important to establish and study the correlation of
$\Delta\vev{p_T^2}$ with $x_B$ and $Q^2$, as emphasized in the
previous section.

\section{Conclusions}

In this paper, I attempted a semi-quantitative estimate of the
prehadron production time, based on recent preliminary \hermes\ data on
hadron $p_T$-broadening in nuclear DIS, see
Eq.~\eqref{eq:prodtimeest}. The obtained production time can well
explain the dependence of the $p_T$-broadening on $A$, $z$ 
and $\nu$. Its linear increase with $Q^2$ remains a challenge to
current theoretical models, and I proposed a few mechanisms that may
contribute to its explanation, also suggesting how to experimentally
validate them. Furthermore, I discussed how the preliminary
\clas\ data qualitatively support the features observed at \hermes.

With the obtained estimate of $t_p$, 
the prehadrons at \hermes\ 
are typically formed around the nuclear surface or
slightly outside, while at \clas\ they are typically formed inside the
target.  However, a quantitative analysis indicate that
$t_p|_\text{\hermes} \approx 0.2 t_p|_\text{\clas}$.
The observed $x_B$ dependence of the $p_T$-broadening
can at least in part reconcile the 2 data sets, as well as explain
the $Q^2$ dependence of the data. It also points at a non-negligible
role of NLO processes in the hadronization process.

The measurement of hadron $p_T$-broadening, and the
related Cronin effect, are beginning to test the limits of current
theory models on the space-time evolution of hadronization, which are
based on LO $\gamma^*$-quark scattering, followed by color
neutralization of the quark and hadron formation.
The main theoretical challenges raised by these data are, in
my opinion, 
\begin{itemize}
\item 
  the inclusion of NLO processes in the modeling of hadron quenching.
\item 
  the implementation at LO, and subsequently at NLO, 
  of alternative models of energy loss, like a medium-modified DGLAP
  evolution;  
\item 
  testing of alternative space-time pictures beyond the 2 time scale
  models currently accepted; for example, the role of a colored
  prehadron should be explored. 
\end{itemize}
On the experimental side, the proposed physics mechanisms can be
verified by measuring the $p_T$-broadening and the Cronin effect with 
suitable kinematic cuts. In particular,
\begin{itemize}
\item 
  the role of parton energy loss can be highlighted by
  large-$\nu$/small-$z$ cuts, such that the parton lifetime exceeds
  the nuclear size;  
\item 
  in-medium hadronization can be selected by small-$\nu$/large-$z$
  cuts;   
\item 
  the $x_B$ dependence of $\Delta\vev{p_T^2}$, the Cronin effect, and
  the multiplicity ratio needs to be better studied over a large
  interval of $x_B$, including small $x_B\lesssim0.01$ and large
  $x_B\gtrsim0.4$; 
  additionally, dihadron correlations and the hadron multiplicity per
  event as a 
  function of $x_B$ can more directly reveal the role of NLO
  processes, in which 2 partons can be produced at the hard
  scattering;
\item
  the combined analysis of $p_T$-broadening's 
  $x_B$-distributions binned in $Q^2$ and of
  $Q^2$-distributions binned in $x_B$ is likely to usefully expose the
  underlying dynamics.
\end{itemize}
Because of the available $\nu$ range, the
\hermes\ experiment and the future Electron-Ion Collider are best
suited to study the role of parton energy
loss and propagation in cold nuclear matter.
At CLAS, in-medium hadronization is likely to be dominant and can be
studied in detail thanks to its high-statistics data, which allow
multidimensional binning and the study of one kinematic variable at a
time.  


\begin{acknowledgments}
I would like to thank E.~Aschenauer, W.~Brooks, J.~Morf\'\i n  for
interesting and informative discussions. I am grateful to P.~di~Nezza
and Y.~Van~Haarlem for additionally and carefully reading the manuscript.
This work has been supported by the DOE contract No. DE-AC05-06OR23177,
under which Jefferson Science Associates, LLC operates Jefferson Lab,
and NSF award No. 0653508.
\end{acknowledgments}


\end{document}